# Adaptive Quorum-based Channel-hopping Distributed Coordination Scheme for Cognitive Radio Networks

Esraa Al Jarrah, Haythem Bany Salameh, Ali Eyadeh

Dept. of Telecommunication Engineering, Yarmouk University, Jordan

*Abstract*—in this paper, we propose novel channel-hopping-based distributed rendezvous algorithm based on grid-based-quorum techniques. The proposed algorithm increases the probability of rendezvous (RDV) within a single cycle by allowing CR nodes to meet more often according to intersection property of quorum systems (QSs). Our proposed algorithm is called Adaptive quorum-based channel-hopping distributed coordination scheme for cognitive radio networks. The main idea of our algorithm is to dynamically adjust the selected QS by CR users according to the varying traffic loads in the CRN. The proposed algorithm decreases the average time to rendezvous (TTR) and increase the probability of RDV. We evaluate the performance of our algorithm through simulations. The performance of our algorithm is compared with two different schemes. The results show that our algorithm can reduce TTR, increase the RDV, and decrease the energy consumption per successful RDV.

## I. INTRODUCTION

In cognitive radio network (CRN) primary users (PUs), when they are not transmitting, introduce spectrum holes. Unlicensed users, which is also called secondary users (SUs) temporally used spectrum holes but without affecting to PUs. They also must vacate the chosen channel whenever PU signals are detected. According to the opportunistic channel access mechanism, SUs must dynamically sense the spectrum holes. If a spectrum hole is not occupied by a PU, it can be used by SUs to establish communication links [1]. When two SUs wish to communicate with each other in CR networks, they must first meet on an unoccupied common channel to exchange control messages. This process is called a "rendezvous". A rendezvous process is an essential part in CRN operation.

Before rendezvous process, users are even not aware of the existence of each other or the different number of available channels of users which is dynamically changing. This makes the implementation of the rendezvous process a challenging problem. Most of existing studies relay on the existence of a dedicated common control channel, which facilities the rendezvous process ([2], [3], and [4]). However, the presence of a dedicated common control channel suffers from two main problems. The first one is the bottleneck problem when there are a large number of packets that the SUs want to transmit through the common control channel, which can cause a high packet collision and reduce the efficiency of channel utilization. In addition, it is very difficult to maintain a single dedicated common control channel due to the dynamic nature of PUs activities and spectrum heterogeneity. There is an alternative method of implementing the rendezvous without the need of a dedicated CCC, which is using channel hopping protocol [5]. Recently, several quorum-based channel hopping schemes (e.g., [5], [6]) have been proposed to overcome the rendezvous (RDV) problem. The concept of quorum has been used to achieve the distribution mutual exclusion, and widely used for a consistent data replication and agreement problems ([7], [8]). Power-saving protocols are considered the main application that use the quorum-based schemes ([9], [10], [11]).

## II. SYSTEM MODEL

### A. CR Network Model

In this paper, we consider a single-hop CRN that coexists with a number of PUs, where all the SUs are within the same coverage area and have the same set of channels. We assume that the CRN contains N licensed channels where $(i = 1,2,...N)$, we also assume that the time T is divided into equal time slots t, where each time slot is enough to exchange the needed control messages. When two nodes want to rendezvous, they can follow 802.11 3-way handshake (RTS/CTS/DTS) process. We note here that MAC design for CRNs is a critical issue, but outside the scope of this paper. We also assume that all nodes in the network are synchronized, where in CRN it is not easy to achieve time synchronization between nodes. Fortunately, various time synchronization protocols were proposed in literate for CRNs [12], [13]. Each SU in the CRN has two half-duplex transceivers, one for control exchange to dynamically switch between the N PR channels and the other one for data transmissions.

### B. The PR Activity Model

In each time slot, we assume the activity of PUs follows the two-state (ON-OFF) Markov model, where the ON period indicates the presence of the PU (PR channel is busy) and the OFF period indicates that the channel is idle. The communication between the PUs is synchronized and the SUs which are also synchronized with the PUs can opportunistically access the available channels. Figure 1 shows the channel state for the ith channel. The PR activities indicate the distribution of the ON-OFF state, where in this model the

ON-OFF states are independent random variables for a given channel. The average idle and busy periods for a given channel i are $\alpha_i$ and $\beta_i$, respectively where ($1 \leq i \leq N$), (e.g., Figures 1 and 2). Then, the idle and busy probabilities for the channel i are given by $P_I = \frac{\alpha_i}{\beta_i + \alpha_i}, 1 \leq i \leq N$, $P_B = 1 - P_I$.

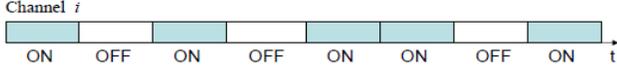

Fig.1: The channel occupancy for the ith channel by PUs [14].

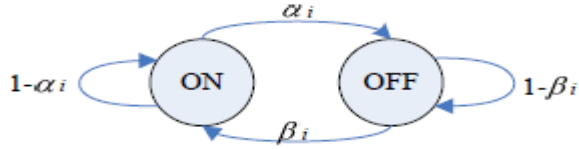

Fig.2: ON-OFF state channel availability model of given PR channel [14].

### III. QOURUM SYSTEM

Quorum systems were primarily developed (and vastly used) in the field of operating systems [15]. Recently, wireless communications also have used quorum systems. The concept of a quorum has been used in achieving the distribution mutual exclusion, and widely used for a consistent data replication and agreement problem [9]. A quorum system has some properties (like intersection property) that can be utilized to establish communications without the need of a CCC, and hence overcome the RDV problem in CRNs. For more information on quorum, we refer the reader to the works in [8], [11]. There are several types of quorum systems such as tree-based [16], grid-based [9], and others [17, 18]. One of these types, which were proposed by Maekawa [9], is the grid-based system. In this system, a grid is used (in the shape of a square matrix) to logically organize the sites (elements) as $n \times n$ array. The union of full row and full column for a requesting site (elements in the array) is a quorum. This system was vastly utilized in power-saving (PS) protocols for wireless networks. In grid-based quorum, the beacon intervals of PS nodes can be divided into groups, which include $n^2$ consecutive intervals, called a quorum group. In this system, each group of $n^2$ interval is organized in $n \times n$ grid matrix, where $n$ is a global parameter. From this grid, one row and one column are arbitrary picked for each node. In [10], the authors defined rules to form a legal grid that called grid allocation rules. To keep the intersection property and guarantee that there are at least two intersections between any two SUs, we cannot arrange the grid randomly.

### IV. THE GRID-BASED QUORUM ALGORITHM

In this paper, we propose a distributed rendezvous algorithm that guarantees RDV during a single period, where the period (cycle) contains a fixed number of a fixed-size time slots. Each cycle continuous $n^2$ time slots (called a group) and each group is arranged in $n \times n$ grid. Each time slot is mapped into one channel. So, the quorum system is used to guarantee RDV by determining which channel is mapped to each time slot. The quorum must be known for all users in the CRN. As mentioned before, we assume that all SUs are within the same coverage area and have the same number of channels. In this algorithm, the grid size depends on the number of all PR channels where $n = \lceil \sqrt{N} \rceil$. Then we map each channel to a grid index, where channel (1) is mapped to index 1, channel (2) is mapped to index 2, channel (3) is mapped to index (3), etc. All the nodes in the CRN have a complete knowledge of the index for each channel and for each user, where each channel will be mapped to one time slot.

In this algorithm, each sender and receiver select a number of rows and columns from the same grid quorum system to generate their channel hopping sequence separately. Thus, the rendezvous will be achieved within a sequence period on their common channels due to the intersection property of quorum system. Consider two users that want to communicate, say users A and B. Recall that all the users in the CRN share the same quorum system. Assume that there are 16 channels in the network, where they are arranged in the grid quorum system. Each user will randomly select one row and one column from the grid-quorum system. The union of the row and column determines the quorum interval, where the users can exchange control information as shown in Figure 3. It is clear that users will intersect on at least two quorum slots. For example, users A and B will exchange control information in channels 7 and 10 and at the 7th and 10th quorum slots.

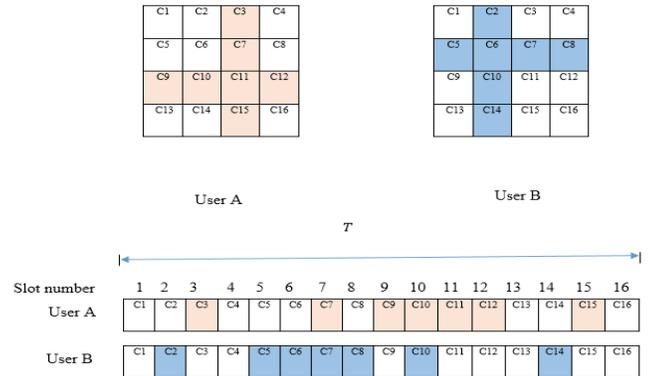

Fig.3: Grid-based Quorum with 16 channels.

### V. THE PROPOSED DESIGN VARIANTS

We propose three different algorithms to choose the rows and columns in the quorum system.

*A. The Proposed 1×1-Qourum RDV Scheme:*
The first algorithm is based on random selection, where each SU randomly selects one row and one column (1×1-selection) from the quorum system and this results in 2n-1 quorum intervals. It is worth mentions that this variant was previously proposed in [19, 20] with different channel mapping procedure. Given two CR nodes, which are perfectly synchronized, we can show that there will be at least two rendezvous in two channels in each cycle.

*B. The Proposed Fixed N×M-Quorum RDV Scheme:*
The second algorithm is based on selecting variable number of rows and columns. In this method, each user determines the appropriate number of rows and columns that a user can select such that the quorum conditions are met. In this method, we study two states. The first one is when users select 2 rows and 1 column (2×1-selection) and the second one, represents the case when users select 2 rows and 2 columns (2×2-selection).

*C. The Proposed Adaptive N×M-Quorum RDV Scheme:*
We also propose an adaptive algorithm that selects dynamic number of rows and columns, where the selection of rows and columns will be adaptively based on the traffic load. We also enhance the previous algorithms by utilizing intersection proprieties in quorum system. We divide the traffic into three regions (low, moderate and high), where for each region we select different number of rows and columns. For low region, we adopt the 2×2-selection which consumes less energy with high average number of successful RDV. The best selection for the moderate region is the 2×1-selection, while for the high traffic region, we use the 1×1-selection, which provides comparable values of average number of successful RDV as the 2×2-selection with less amount of energy consumption.

➢ Practical implementation

We assume the traffic estimation mechanism in [21] is in place to determine the various traffic regions (low, moderate and high) the details of such mechanism are outside the scope of our work and left for future work.

## VI. PERFORMANCE EVALUATION

*A. Simulation Setup:*
We analyze the proposed schemes using MATLAB simulations. In our simulations, we consider that there are several PRNs which coexist with a CRN. We test the performance of our schemes for (N=16) channels and for grid size $n \times n$. We consider that there are 26 or 50 transmitting CR users. At each time slot, the CR transmitter randomly selects its receiver. Each node can perform traffic estimation for the PUs and SUs. We model the availability of channels by using the 2-states Markov chain. Our results are averaged over 800000 time slots.

*B. Simulation Results:*
In our simulations, we use four metrics to compare the performance of our different schemes, i.e.

- The average number of successful rendezvous, which is defined as the average number of successful meeting per time slot [22].
- Average TTR, which is the number of time slots that two users need to wait on average before they found common channel to rendezvous [23].
- Normalized energy per successful RDV, which indicates the average number of active slots per a successful transmission.
- Forced blocking due to PR activity, which indicates the effect of PR activity on network performance [24].

The performance of the proposed variants (2×1-selection, 2×2-selection and adaptive selection) is compared with a reference scheme that uses only 1×1-selection [19], [20].

*1) Average Number of Successful Rendezvous:*
First, we study the average number of RDVs as a function of $P_I$ in Figure 4. We can observe that the selection of 2-rows and 2-columns achieves the best performance. This is expected as its quorum interval is longer than that of the other schemes, and hence the number of intersections between nodes increases. Also, the average number of rendezvous increases as the channel availability increases.

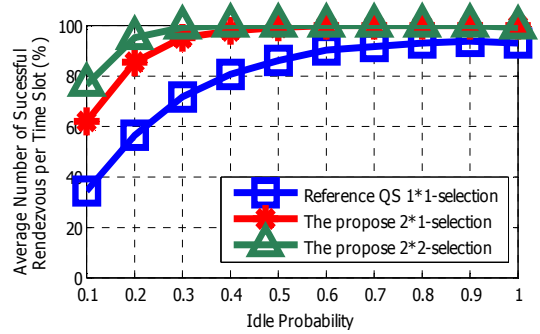

Fig.4: Average number of successful rendezvous per time slot vs. $P_I$.

*2) Average TTR*
Figure 5 shows the average TTR as a function of $P_I$ for 50 CR users. The average TTR decreases as $P_I$ increases due to the fact that when the availability of the channel is high, SUs pairs achieve RDV without encountering PU effects. Note that the 2×2-selection achieves the best average TTR performance among other two schemes due to the fact that it has more intersection slots.

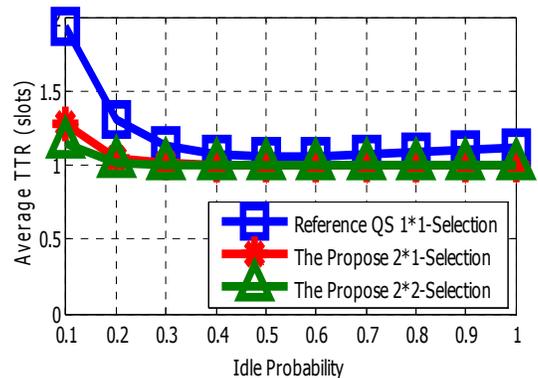

Fig.5: Average TTR (slots) vs. $P_I$ (SUs=50).

### 3) Normalized Energy per Successful RDV

Next, we study the normalized energy consumption per a successful RDV, which indicates the number of units of energy consumed per a successful exchange assuming that each time slot consumes 1 unit of energy:

$$\text{Normalized energy per a successful RDV} = (1 \text{ unit of energy}) \frac{\text{Total number of active slots}}{\text{Total number of RDV}}$$

Figure 6 shows the normalized energy per a successful RDV as a function of $P_I$.

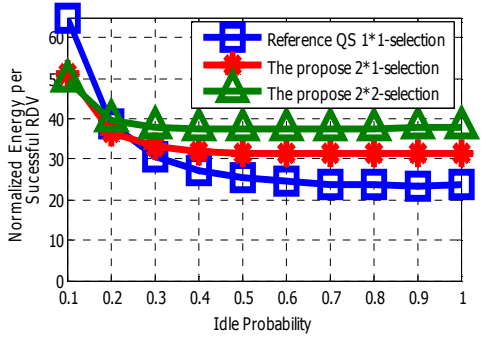

Fig.6: Normalized energy per successful RDV vs $P_I$ (SU=50).

### 4) Forced Blocking

The forced blocking probability of SUs can be expressed as the average number of blocking per time slot of SUs divided by the average number of successful meetings per time slot:

$$\text{Forced blocking probability} = \frac{\text{Average number of blocking}}{\text{Average number of successful meeting}}$$

Figure 7 shows that similar behavior is observed for all schemes. Note that as $P_I$ increases, the forced blocking probability decreases. This is because at low $P_I$ the number of channels will not be sufficient to support any SU transmission and so new SUs will be blocked. At high $P_I$, the number of available channels increases, so the communications between SUs will also increase. Thus, we conclude that the forced-blocking depends on channel-availability and PR activity.

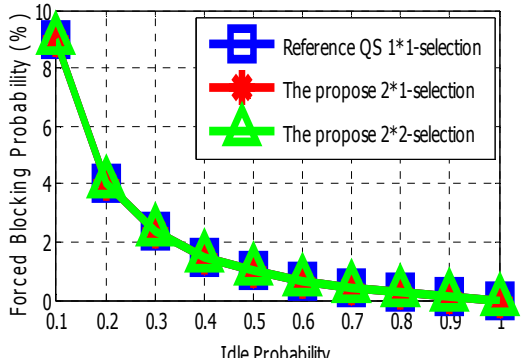

Fig.7: Forced blocking vs. $P_I$ (SUs=50).

### 5) Performance Evaluation of Adaptive Algorithm

We need an adaptive approach to improve the performance in terms of number of RDV, average TTR and the number of units of consumed energy per successful RDV, so it's necessary to allow different CR users to dynamically select appropriate rows and columns from the quorum system based on the traffic load and network performance requirements. From the previous results, we observed that the 2×2-selection always gives the best performance of number of successful RDV but consumed more energy, 1×1-selection consumed less energy and achieved less number of successful RDV and 2×1-selection consumed less energy than the 2×2-selection and achieved more number of successful RDV than the 1×1-selection.

For each metric, we evaluated the adaptive algorithm for different number of users to achieve best performance of average number of successful RDV with less energy consumption and study the effect of varying the availability on channel on different number of users for grid size 4×4. Figure 8 shows the average number of rendezvous performance.

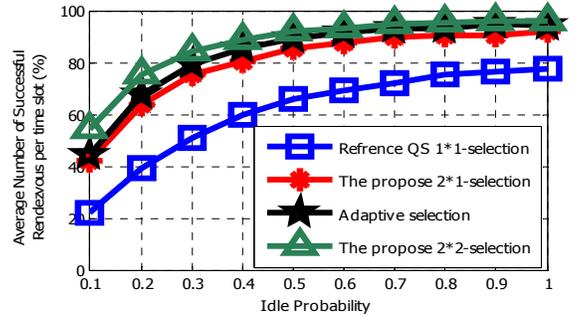

Fig.8: Performance evaluation: average number of successful rendezvous per time slot vs. $P_I$.

We study the average TTR and the normalized energy consumption per successful RDV in Figures 9 and 10, respectively. The results show that our algorithm decreases the energy consumption per successful RDV.

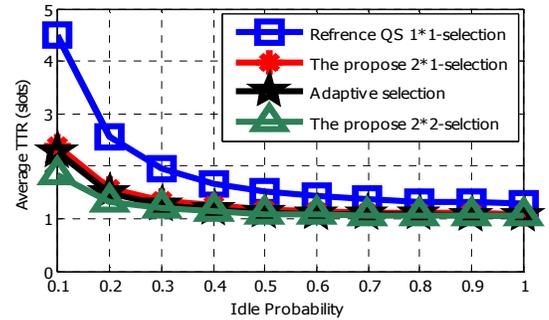

Fig.9: Performance evaluation: average TTR (slots) vs. $P_I$.

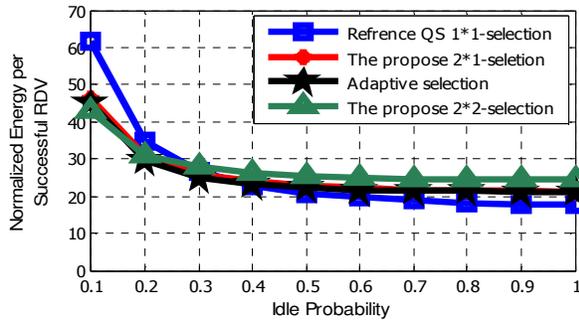

Fig.10: Performance evaluation: normalized energy per successful RDV vs. $P_I$.

VII. CONCLUSION

In this paper, we have studied the grid quorum system to develop new channel-hopping-based distributed rendezvous algorithms. We proposed and analyzed three different channel selection schemes. We solve the RDV problem by increasing the number of common channels between any pair of CH sequence. The performance analysis verified that there is tradeoff between the four metrics for different selection schemes. In summary, we need an adaptive approach, which dynamically adapts the selection of the number of rows and columns to achieve the best required performance for our network based on the main objective (i.e., the average number of RDVs or the amount of energy consumption per a successful RDV).